**Large-angle, gigahertz-rate random telegraph switching induced by spin-momentum transfer**

Matthew R. Pufall, William H. Rippard, Shehzaad Kaka,

Steven E. Russek, Thomas J. Silva

*Electromagnetics Division 818, National Institute of Standards and Technology,*

*Boulder, CO 80305*

Jordan Katine, Matt Carey

*Hitachi Global Storage Technologies, San Jose, CA 95120*

**Abstract:**

We show that spin-polarized dc current passing through a small magnetic element induces two-state, random telegraph switching of the magnetization via the spin-momentum transfer effect. The resistances of the states differ by up to 50% of the change due to complete magnetization reversal. Fluctuations are seen for a wide range of currents and magnetic fields, with rates that can exceed 2 GHz, and involve collective motion of a large volume ($10^4$ $nm^3$) of spins. Switching rate trends with field and current indicate that increasing temperature alone cannot explain the dynamics. The rates approach a stochastic regime wherein dynamics are governed by both precessional motion and thermal perturbations.

The recent observations of high-frequency precessional motion and magnetization switching induced by a spin-polarized dc current have further spurred the study of the spin momentum transfer (SMT) effect as a possible means of efficiently switching the small magnetic elements in magnetic memory (MRAM), and as the basis for high-frequency, nanometer-sized microwave oscillators.[1,2,3,4] In addition to high-frequency precessional motion, the excitation of an unexpected lower-frequency "broadband instability" extending from dc to several gigahertz has also been observed.[1,3] Here, we report real-time measurements of this SMT-induced lower-frequency response of spin-valve structures, both of patterned structures and of lithographic point contacts made to continuous films. We show that the broadband instability is the result of the magnetization in these structures undergoing large-angle fluctuations between two distinct states in response to the dc current. These fluctuations have the characteristics of classic random telegraph switching (RTS), and are observed for a wide range of currents and applied magnetic fields, and for various device geometries and sizes. The fluctuations occur on time scales ranging from microseconds to fractions of a nanosecond, and can be observed over the same range of currents and fields as the coherent high-frequency precessional excitations reported previously.[1,3] As the dimensions of magnetoelectronic devices decrease, as for hard-disk drive read heads and spintronic devices, device volumes will reach the point where SMT effects such as those reported here must be addressed.

Two-state systems exhibiting RTS—that is, each state having an independent dwell time—have instances across the physical sciences, from solid state physics[5,6,7,8,9] to biology[10], as well as in magnetic systems.[11,12,13] Spin-transfer-induced RTS at Hertz to KHz rates has also been observed previously[14] for systems in which two well-defined magnetostatic states are accessible, i.e. when the applied field $H_{\mathrm{app}}$ is less than $H_c$, the coercive field of a patterned device with uniaxial anisotropy. The spin-transfer current modifies the thermally-activated fluctuation rate between the two states. Here we explore spin-transfer-induced effects for larger applied fields $H_{app} > H_c$, so that the applied field is of a magnitude sufficient to allow only one stable state in the absence of an applied current. It has been previously shown that SMT can induce switching[15,16] of the magnetization between parallel and anti-parallel (relative to the applied field) orientations, at rates < 1 MHz. The observations reported here significantly extend these measurements, showing that this spin-transfer-induced random telegraph switching persists in higher fields, and out to very high rates (in excess of 1 GHz). Consequently, these systems approach a dynamical regime in which the switching rate approaches the intrinsic damping rate of **M**, and the dynamics of **M** have both precessional and thermal characteristics. Furthermore, we show that even at these rates, the fluctuation amplitude is large but no longer corresponds to $180^o$ reversal of the magnetization. Instead, the device exhibits resistance changes of up to 50% of complete reversal, indicating that the spin-torque induces a second (meta)stable configuration of the magnetization, the properties of which are a function of both applied field and current.

These features distinguish spin-transfer-induced telegraph switching relative to other two-state systems: RTS in other physical systems frequently results from the fluctuation of a small defect or region of a larger structure, typically showing characteristic rates from sub-hertz to megahertz before the breakdown of two-state behavior. In contrast, SMT-induced two-state switching corresponds to the large-amplitude *collective* motion of the entire device, a volume of $\approx 10^4$ nm$^3$, at GHz rates. Though the system is driven by the spin transfer current, we show that nonetheless the general analytical techniques developed for the study of two-state switching systems enable us to study in new ways the interaction between the polarized current and the local magnetic moment, and we show evidence that SMT significantly modifies the energy surface experienced by the free layer magnetization.

Room temperature, high-bandwidth *I-V* resistance measurements were made on Cu (100 nm)/IrMn (7 nm)/Co (7.5 nm)/Cu (4 nm)/Co (3 nm)/ Cu (20 nm)/ Au (150 nm) spin-valve structures patterned into 50 nm x 100 nm elongated hexagonal pillars. The exchange bias field from the IrMn layer (<1 mT) did not have a significant effect on the magnetics. The thinner Co layer, having a lower total moment, responds more readily to spin torques than the thick Co layer.[4] These layers are referred to as the "free" magnetization ($\mathbf{M}_{\text{free}}$) and the "fixed" magnetization ($\mathbf{M}_{\text{fixed}}$) layers, respectively (see Fig. 1a inset). In a spin-valve, changes in the relative alignment of the magnetizations of these two layers result in variations in device resistance through the giant magnetoresistance (GMR) effect. By dc current-biasing the device, these resistance changes are seen as changes in the voltage across the device. In the following analysis, we assume that all SMT-induced voltage changes in these devices are due to GMR.[17,18] Measurements were also made on larger (75 nm x150 nm) elongated hexagons, circular devices of 50-100 nm radii, and lithographically-defined 40 nm point contacts made to unpatterned spin-valve multilayers.[3] All devices showed similar RTS, indicating that the bistability is a generic feature associated with SMT rather than an artifact of a specific device geometry.

When a dc current $I$ is driven through these devices in the presence of an in-plane field $H_{\text{app}}$ sufficient to align the magnetizations of the two layers, we observe a reversible step in the dc resistance at a critical current $I_c$, as has been described elsewhere.[3,4,19] A typical dc resistance trace is shown in Fig. 1a, with $I_c \approx 6$ mA (corresponding to a current density $J_c \approx 10^8$ A/cm$^2$). This step occurs for only one sign of current, with electrons flowing from the free to fixed layer, indicating that the resistance change is due to spin-torque rather than current-generated magnetic fields (see Fig 1a inset).[4,20] The position of the step is a function of applied field, and is also, as shown in the real-time voltage traces in Fig. 1b, correlated as follows with the appearance of two-state telegraph switching of $\mathbf{M}_{\text{free}}$.[15] Below $I_c$, the device lies primarily in the low resistance state, switching occasionally but rapidly into (and out of) the higher resistance state. As the current approaches $I_c$ (see Fig. 1b), this switching occurs with greater frequency. The transient switching time between the states is quite fast, and varies with $I$ from 1 ns to 2 ns for low currents, to ≈ 0.7 ns for higher $I$. As $I$ increases, the characteristic time spent in each state changes, with the time spent in $R_{\text{low}}$, the dwell time ($\tau_{\text{low}}$) decreasing, and the time in $R_{\text{high}}$, $\tau_{\text{high}}$, increasing. The step in dc resistance occurs where $\tau_{\text{low}} \approx \tau_{\text{high}}$. Above $I_c$, $R_{\text{high}}$ becomes most likely (a fact also reflected in the dc resistance trace.)

The states between which the magnetization switches are functions of both current and field. From triggered real-time voltage traces taken at a fixed current and field (taken on a different device than for the data shown in Fig.1), we determined the distribution of voltage fluctuations (as it is an ac-coupled measurement), and from this calculated the resistance change $\Delta R$ between the two states. As shown in Fig. 2c, $\Delta R$ varies with current and applied field, and is not monotonic with $I$ for a given $H_{\text{app}}$, having a maximum value around $I_{\text{c}}$. On the other hand, the peak value in $\Delta R$ decreases approximately linearly with $H_{\text{app}}$, in a manner similar to the dc resistance step (see Fig 2b).

Field- (see Fig. 2a) and current-induced switching measurements[4] indicate that these devices are sufficiently small such that the free layer magnetization vector $\mathbf{M}_{\text{free}}$ behaves as a single domain, and that $|\mathbf{M}_{\text{free}}|$ is not reduced by the current.[21] Assuming this, a cosine dependence of $R(\theta)$ due to GMR, and taking one of the states as $\mathbf{M}_{\text{free}} \parallel \mathbf{M}_{\text{fixed}}$ (because $\Delta R$ and angle are not uniquely related), we estimated the dynamic rotation angle of the free layer, shown on the right axis in Fig. 2c. The motions of $\mathbf{M}_{\text{free}}$ induced by the dc current are not simply small perturbations about equilibrium, as is typically observed in thermally driven two-state magnetic systems[7,11,13]: For low fields and currents, the magnetization rotates a full $90^o$, a value that decreases to $\approx 40^o$ at higher $H_{app}$ and $I$. The larger amplitudes previously observed at lower fields and fluctuation rates[15] are consistent with this trend.

The characteristic times $\tau_{high}$ and $\tau_{low}$ at a given $I$ and $H_{app}$ were determined from the conditional probability distribution functions $P(R_{high}|R_{high}\ @\ t = 0)$ and $P(R_{low}|R_{low}\ @\ t = 0)$. These probabilities were constructed from multiple triggered real-time traces, and show an exponential falloff with time. Fitting the expression $P(t) = e^{-t/\tau}$ (in which $t$ is the elapsed time after a switch, and $\tau$ is $\tau_{high,low}$) to the measured distributions gives $\tau_{high}$ and $\tau_{low}$, shown as functions of $I$ for several fields in Figs. 3(a-f). Measurements were made from 4 mA to 12 mA. Dwell times longer than the acquisition wait time ($\approx$1 ms before auto-triggering) were not recordable; consequently very long dwell times (i.e., quasi-stable magnetic configurations) are indicated by an absence of dwell times at that current. The $\tau$ of each state varies by several orders magnitude over the applied current range, corresponding to states that vary from nominally long-lived ($\tau_{dwell} > 8$ μs) to highly unstable ($\tau_{dwell} \approx 1$ ns). The plots in Fig. 3 represent quantitatively what was described qualitatively for Fig. 1: Initially $\tau_{low} >> \tau_{high}$, and as $I$ increases, the dwell times converge, cross at $I_c$ (at the step in dc resistance[15]), and diverge, with $\tau_{high}$ becoming the longer dwell time.

As seen in Figs. 3(a-f), $\tau_{high}$ varies with current in a markedly different fashion from $\tau_{low}$. On a logarithmic scale, $\tau_{low}$ is roughly linear with current, whereas $\tau_{high}$ is a more complicated function of current. Furthermore, the functional form of $\tau_{high}$ is a strong function of applied field, whereas $\tau_{low}$ remains roughly linear, primarily shifting to higher currents and changing slope slightly. The states, and the potential barrier between them inhibiting fluctuations, are evidently distinct functions of current and field. Finally, the range of currents over which fluctuations are observed increases with field: The onset current for fluctuations is a relatively weak function of field, while the current at which fluctuations cease moves rapidly out to higher currents. So, though more field is being applied, the fluctuations are not more stabilized.

For $\mu_0 H_{app} > 0.19$ T, the instability is still observed, but the fluctuation rates exceed the bandwidth of the real-time oscilloscope. However, with some general assumptions about the system, the dwell times can instead be determined from the power spectra. Two-state or telegraph noise, in which the dwell times in the two states are independent, produces a Lorentzian power spectral density, centered about dc[9], of the form

$$S(\omega) = \frac{\Delta V^2}{4\pi} \frac{\tau_h \tau_l}{\tau_{eff}^2} \frac{1/\tau_{eff}}{\omega^2 + 1/\tau_{eff}^2}, \quad (1)$$

in which $\Delta V = I\Delta R$ is the change in voltage, and $1/\tau_{eff} = 1/\tau_{high} + 1/\tau_{low}$. Power spectra for several currents with $\mu_0 H_{app} = 0.24$ T are shown in Fig 3g. Lorentzians fit the data well, as shown for $I = 12$ mA. From the width, a $\tau_{eff}$ of 0.37 ns is determined, corresponding to a single-state switching rate in the range 1.4 GHz - 2.7 GHz.[22] In other devices, we observed rates in excess of 4 GHz. The lack of deviations of the spectral shape from a dc-centered Lorentzian indicates the validity of the random, two-state assumption. This signature of telegraph switching was observed for a wide range of fields: Even fields in excess of 0.5 T were insufficient to suppress the SMT-induced RTS.

It is important to note that these SMT-induced two-state fluctuations are observed for applied fields such that, *magnetostatically,* only one orientation of $\mathbf{M}_{free}$ is allowed: In the absence of current, no second state into which the magnetization can switch, and the potential surface is parabolic with a single mininum. The spin-polarized current *induces* a second meta-stable magnetization configuration at sufficiently high current densities, analogous to an applied field. This is a somewhat surprising conclusion, since typical formulations write spin-torque as an "anti-damping" term in a Landau-Lifshitz Gilbert dynamical equation, rather than as a field, and damping will ordinarily not affect the energy states of a system. Indeed, while single-domain numerical models[1,23] based on Slonczewski's theory of SMT-induced magnetization dynamics[20] have reproduced aspects of the precessional motion observed in Refs. 1 and 3, these models fail to describe this broadband instability. Stochastic switching has been observed in simulations, but either only over small current ranges near transitions between two different current-dependent precessional states (rather than over a large range of $I$ as shown here), or between two states with identical GMR values.[24] In addition, these simulations describe fluctuations between precessional states. However, we have observed two-state switching both in the presence of and *independent* of the existence of high-frequency precessional signals, up to the bandwidth limit of our system (≈12 GHz, for the nanopillar structures, 40 GHz for the lithographic point contacts).

The random occurrences of the two-state fluctuations such as those shown in Fig. 1b certainly suggest an Arrhenius-Néel two-state analysis describing a thermally activated process over a barrier as a way of parameterizing the effects of the spin-torque and the applied field on the system. Previous work also used an Arrhenius-type analysis to parameterize the low-frequency RTS, employing an independent magnetic temperature for each state.[15] This work used a two-temperature model that does not succeed in describing our data over the ranges of currents and fields presented here[25]; here we will use a single temperature model. A possible complicating factor against using an Arrhenius-Néel analysis (a factor that also drives new effects such as the observed high-frequency precession[3]), is that torques due to SMT are not conservative, and as a consequence the work done by the torque is path-dependent.[26] Therefore, it is not necessarily the case that a "barrier" is well-defined for this system. Nonetheless, previous experimental[14] and theoretical[26] work supports the applicability of a two-state model[27] if the energy barrier is replaced by an "activation energy" that includes the work done *by* the spin torque.

Subject to these caveats, the expression

$$\tau_{\text{high, low}} = \tau_0 e^{U(I,H)_{\text{high, low}}/k_\text{B}T} \qquad (2)$$

describes the characteristic time τ for thermal activation over a barrier, in which $U(I,H)_{\text{high,low}}$ are the respective *effective* barrier heights for the high and low resistance states (shown schematically in Fig. 4), $1/\tau_0$ is the "attempt frequency," $k_\text{B}$ is Boltzmann's constant, and $T$ is the magnetic system temperature, which is possibly different from the phonon (system) temperature. By comparing $\ln(\tau_{\text{high}}/\tau_{\text{low}}) = (U_{\text{high}} - U_{\text{low}})/k_\text{B}T$ to $\ln(\tau_{\text{high}}\cdot\tau_{\text{low}}/\tau_0^2) = (U_{\text{high}} + U_{\text{low}})/k_\text{B}T$ we can estimate a *upper* limit on $\tau_0$ of 1 ns[28], of the same order as the dwell time determined from the power spectrum in Fig. 3b. The attempt frequency $1/\tau_0$ is set by the intrinsic damping rate of **M** and is typically several gigahertz for these systems[29], consistent with these values.

Assuming a value of 0.5 ns for $\tau_0$, we calculated the effective barrier heights $U_{\text{high}}/k_\text{B}\underline{T}$ and $U_{\text{low}}/k_\text{B}T$ as a function of $I$, shown for several fields in Fig. 4. These plots show that the current modifies the effective barrier seen by each state from roughly 1 to 10 $k_\text{B}T$ over the measurement range. By comparison, the barrier height between the two bistable states at zero field and current is over 100 times this value. As expected, the barrier heights vary opposite to each other with current. For low currents, the low resistance ($\mathbf{M}_{\text{free}} \parallel \mathbf{M}_{\text{fixed}}$) state is preferred, having a larger barrier impeding escape, while the high resistance state becomes the more probable state at higher currents with $I > I_c$. Recall that this high-resistance state does not exist in the absence of the current for these applied fields, as there is no secondary potential minimum. In Fig 4, we can see evolution of the state, with the current progressively increasing the barrier height defining it. However, though the two barriers have opposite trends with current, the functional forms are markedly different: $U_{\mathrm{low}}$ is roughly a linear function of $I$, while $U_{\mathrm{high}}$ displays a more complicated dependence. In this sense, these results deviate substantially from previous, low-frequency SMT-mediated two-state switching data[14,15] in which both states showed a linear dependence on current (though over a smaller field and current range). In addition, it is also clear that the effective barrier heights do not asymptotically approach zero with increasing current, as would be the case if an Ohmic or other monotonic temperature increase were the source of the telegraph noise. Indeed, temperatures far above the Curie temperature of the ferromagnet would be required to reduce the ratio $U/k_{\mathrm{B}}T$ sufficiently to approach the observed switching rates.

We have shown that a spin-polarized dc current passing through a small magnetic element induces an instability in the magnetization, resulting in classic two-state, random telegraph switching for a wide range of currents and for applied magnetic fields that magnetostatically allow only one orientation of **M**. A large magnetic volume collectively undergoes large-amplitude fluctuations between two states, at gigahertz rates, resulting in a broadband dc-centered Lorentzian power spectrum. These fast fluctuation rates suggest that SMT significantly modifies the potential surface experienced by **M**. The fluctuation rate trends observed with applied field and current indicate that an increase in temperature – whether of the device or magnetic system – does not by itself describe the dynamics. With the rapidly shrinking device dimensions in magnetoelectronics, device volumes will reach the point where SMT effects are unavoidable. The results presented here indicate that attention should be paid to SMT-induced dynamics as a source of broadband, large-amplitude noise.

**Figure Captions**:

Figure 1: (a) dc *I-V* curve for 100 nm x 50 nm spin-valve nanopillar device for $\mu_0 H_{app}$ = 0.1 T. SMT-induced step in resistance denoted by $I_c$. Inset graphic shows device structure, field and current flow directions. Black circles denote currents for real time traces shown in lower panel. (b) Real time AC voltage traces taken at specified currents. $\Delta R$ defined as noted on trace at $I$ = 5.2 mA, individual dwell events shown for $I$ = 5.7 mA. Device dc response and ac-coupled time-dependent response monitored as functions of current and applied field, through 40 GHz probe contacts. Real time traces acquired with a 1.5 GHz (sampling rate of $8\times10^9$/s) bandwidth single-shot real-time oscilloscope, used to either capture traces up to 8 μs in length (as in Fig 1b), or to measure pulse widths and heights of a succession of individual triggered switching events at a fixed current and field, as in subsequent figures. Power spectra (see Fig. 3g) acquired with a 50 GHz spectrum analyzer.

Figure 2: (color online) (a) *R-H* curve of device, showing $\Delta R$ for 180° rotation of layers. (b) dc *I-V* traces for nanopillar, as a function of field, with current scanned both up and down. Note small hysteresis between up and down scans, also manifested in dynamic $\Delta R$ traces and Fig. 3 time traces. (c) Measured dynamic $\Delta R$ for 100 nm x 50 nm nanopillar, as a function of current and $\mu_0 H_{app}$. Device structure is the same as for Fig.1. $\Delta R$ defined as in Fig. 1b. $\Delta R$ determined from 150 triggered switches. Error bars are $\sigma_{\bar{x}}$ from a Gaussian fit to $\Delta R$ distribution at fixed $I$, $\mu_0 H_{app}$. Right axis: estimated angular deviations of $\mathbf{M}_{free}$, using model described in text.

Figure 3: (a-f) Measured dwell times $\tau_{high}$ and $\tau_{low}$ vs. current, for several applied fields $\mu_0 H_{app}$. Dwell times determined from measured conditional probabilities of the form $P_{high(low)}(t) = P(R_{high(low)}|R_{high(low)}\ @\ t = 0)$, constructed from 150 triggered events, as described in text. (g) Power spectrum of device at $\mu_0 H_{app}$ = 0.24 T for several device currents. Lorentzian fit to $I$ = 12 mA shown.

Figure 4: (color online) Plot of calculated effective energy barriers $U_{high}$ and $U_{low}$, in units of $k_B T$, as a function of current for several fields. A value of $\tau_0$ = 0.5 ns was assumed. Data for up and down scans of current averaged together for clarity. Inset: Diagram showing definition of effective energy barriers.

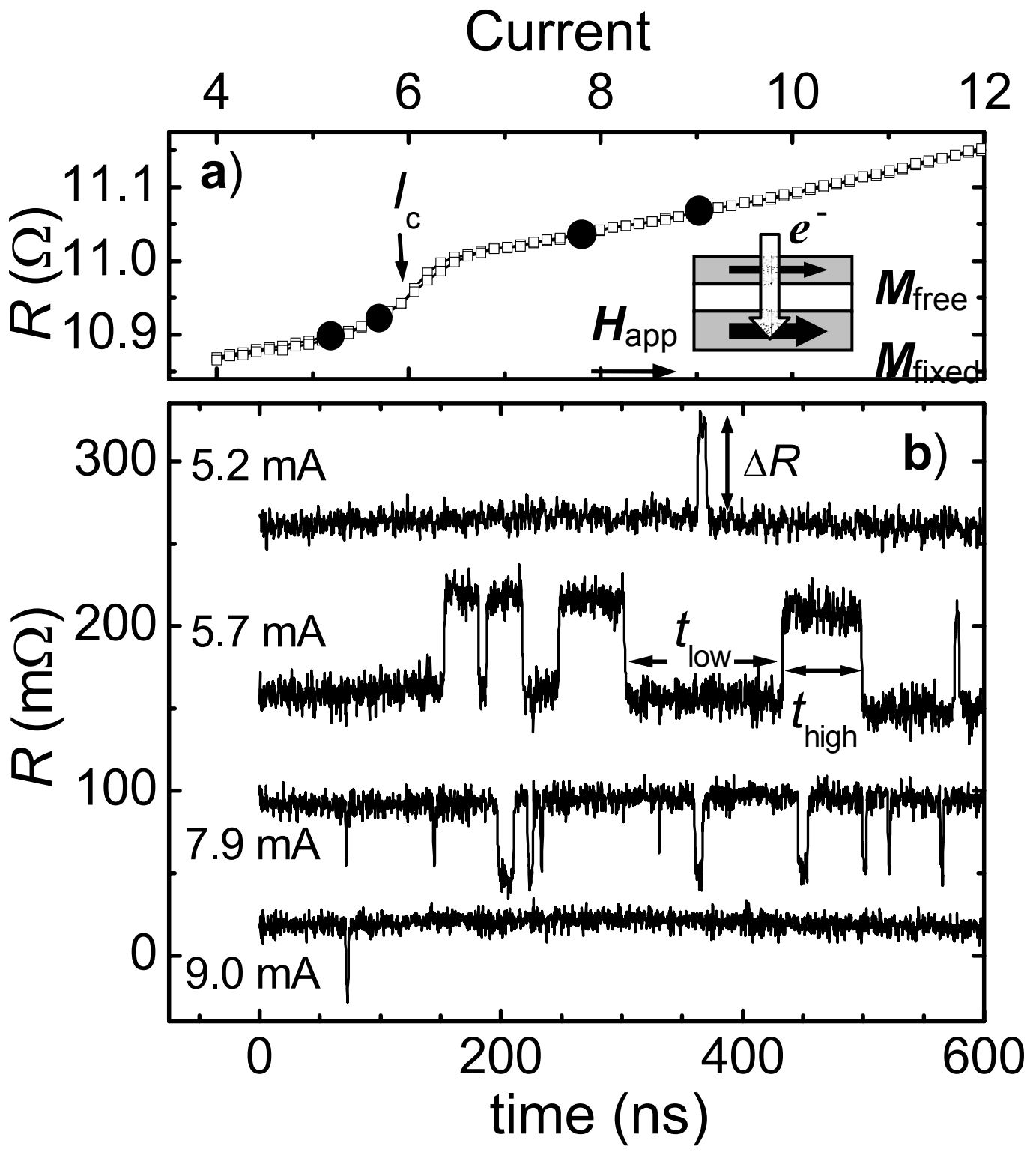

Figure 1.

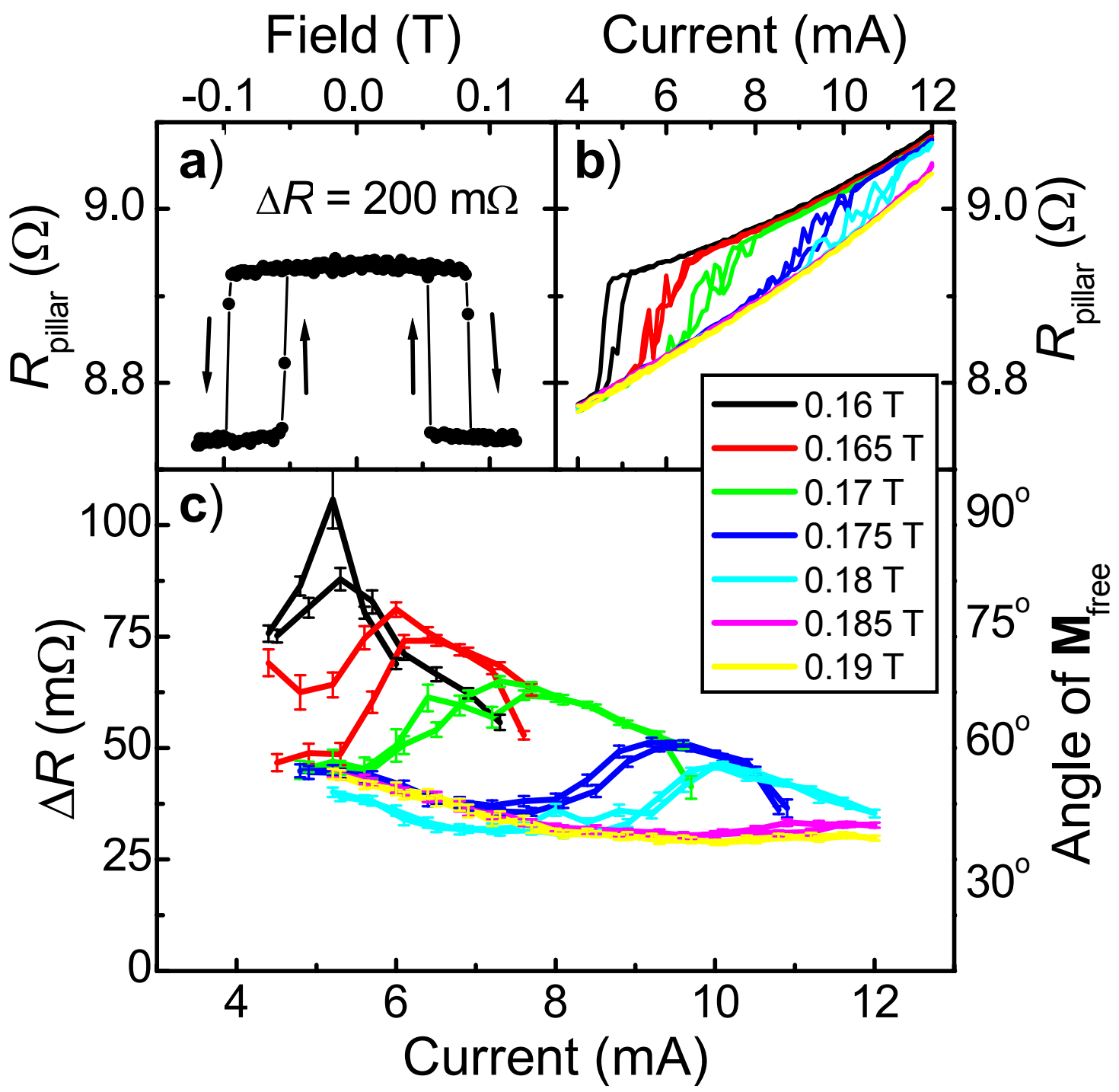

Figure 2.

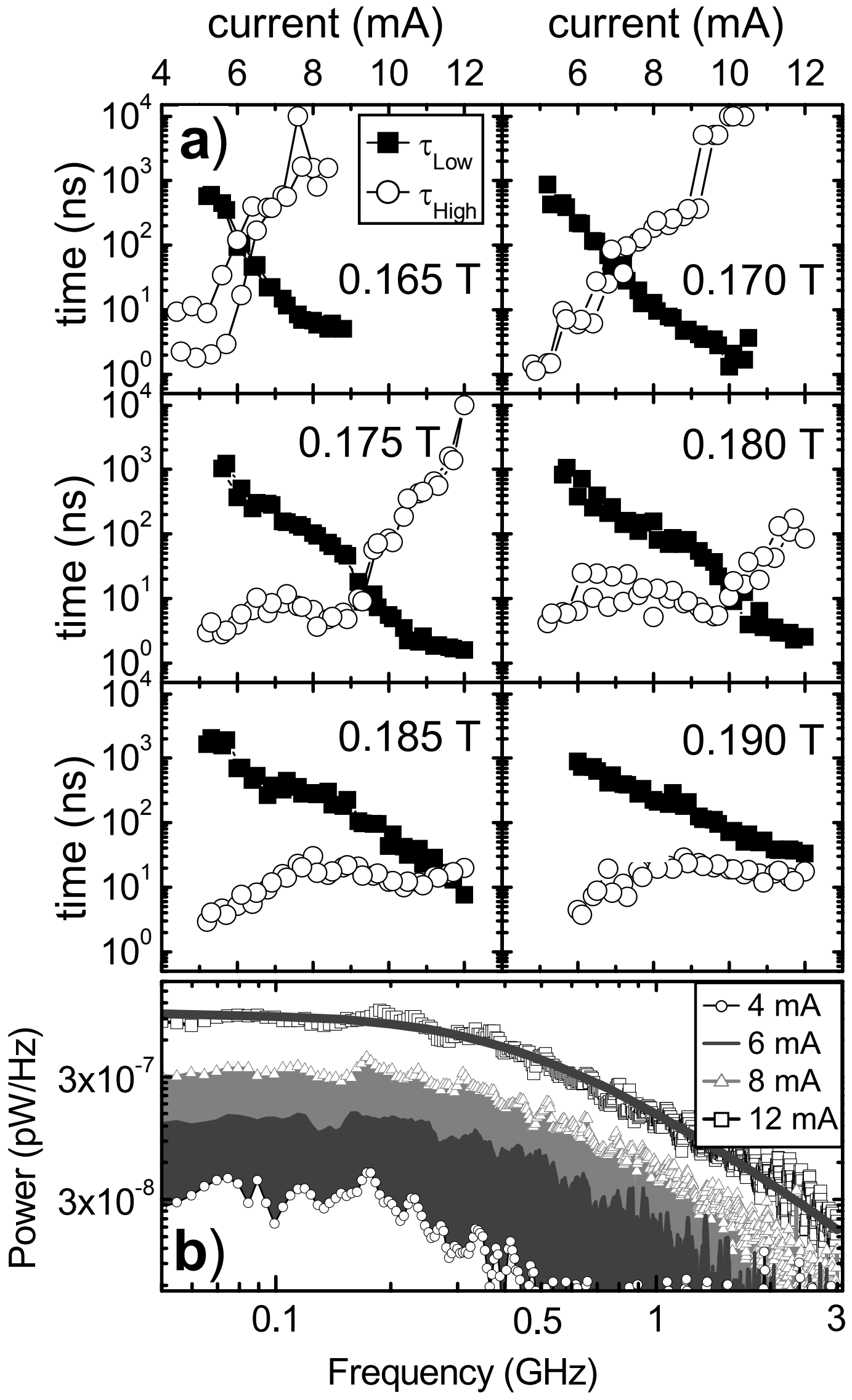

Figure 3.

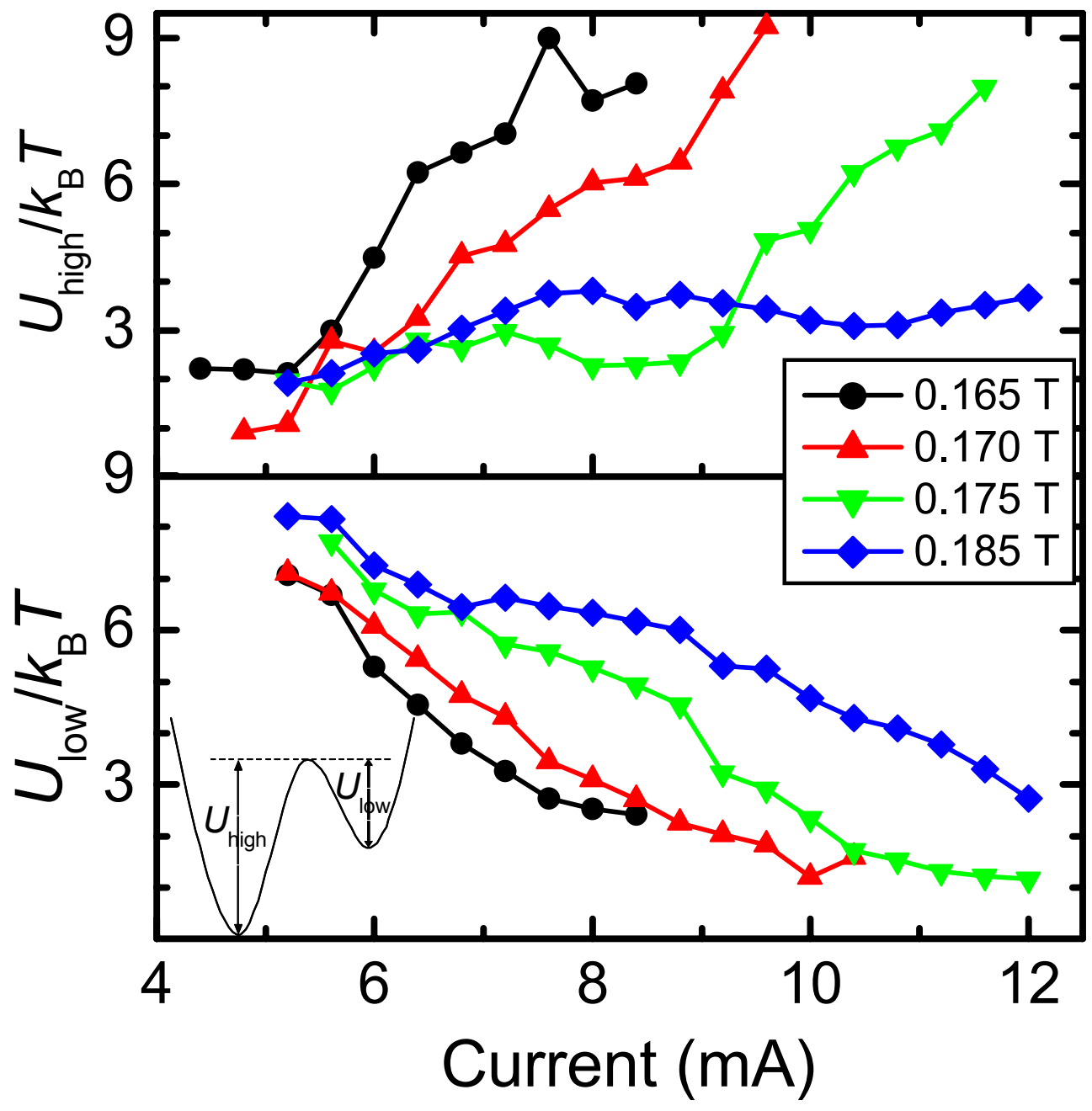

Figure 4.